\documentclass[runningheads]{llncs}
\usepackage{graphicx}
\usepackage{xcolor}
\usepackage{booktabs}%
\usepackage{multicol}
\usepackage{graphicx}
\usepackage{tcolorbox}

\usepackage{amsmath} 

\definecolor{todoRed}{RGB}{255,0,0} 

\usepackage{etoolbox} 
\makeatletter
\patchcmd{\ps@headings}{\rlap{\thepage}}{}{}{}
\patchcmd{\ps@headings}{\llap{\thepage}}{}{}{}
\makeatother
\begin{document}
%
\title{A Cross-Cultural Assessment of Human Ability to Detect LLM-Generated Fake News\\about South Africa}
%
\titlerunning{Human Ability to Detect LLM-Generated Fake News about South Africa}
%

\author{Tim Schlippe\inst{1}\orcidID{0000-0002-9462-8610}, Matthias Wölfel\inst{2}\orcidID{0000-0003-1601-5146}, Koena Ronny Mabokela\inst{3}\orcidID{0000-0002-8058-969X}}
\authorrunning{T. Schlippe and M. Wölfel, K. R. Mabokela}
\institute{IU International University of Applied Sciences, Germany
 \and
 Karlsruhe University of Applied Sciences, Germany
 \and
Applied Information Systems, University of Johannesburg, South Africa\\
\email{tim.schlippe@iu.org}}
\maketitle   

\begin{abstract}
This study investigates how cultural proximity affects the ability to detect AI-generated \textit{fake} news by comparing South African participants with those from other nationalities. As large language models increasingly enable the creation of sophisticated \textit{fake} news, understanding human detection capabilities becomes crucial, particularly across different cultural contexts. We conducted a survey where 89 participants (56~South Africans, 33~from other nationalities) evaluated 10~\textit{true} South African news articles and 10 AI-generated \textit{fake} versions. Results reveal an asymmetric pattern: South Africans demonstrated superior performance in detecting \textit{true} news about their country (40\% deviation from ideal rating) compared to other participants (52\%), but performed worse at identifying \textit{fake} news (62\% vs. 55\%). This difference may reflect South Africans' higher overall trust in news sources. Our analysis further shows that South Africans relied more on content knowledge and contextual understanding when judging credibility, while participants from other countries emphasised formal linguistic features such as grammar and structure. Overall, the deviation from ideal rating was similar between groups (51\% vs. 53\%), suggesting that cultural familiarity appears to aid verification of authentic information but may also introduce bias when evaluating fabricated content. These insights contribute to understanding cross-cultural dimensions of misinformation detection and inform strategies for combating AI-generated \textit{fake} news in increasingly globalised information ecosystems where content crosses cultural and geographical boundaries.

\keywords{Fake News  \and Disinformation \and Natural Language Processing \and South Africa \and Large Language Models \and LLMs.}
\end{abstract}

\section{Introduction}

Large language models (LLMs) have dramatically lowered barriers to producing convincing \textit{fake} news~\cite{zellers2020defending,bommasani2022opportunities}. Unlike traditional misinformation with obvious flaws, AI-generated content is coherent, sophisticated, and increasingly indistinguishable from \textit{true} news~\cite{clark2021all,mitchell2023detectgpt}. This technological shift creates an urgent need to understand how humans detect---or fail to detect---AI-generated \textit{fake} news.

This challenge intensifies when readers lack cultural or factual familiarity with the subject matter, as they struggle to identify inaccuracies that would normally serve as red flags~\cite{guess2020less,vosoughi2018spread}. This cultural dimension of \textit{fake} news detection remains underexplored, particularly for AI-generated content about regions with less global media attention, such as South Africa~\cite{wasserman2020fake,mare2019fake}.

This study investigates the influence of cultural proximity on the identification of AI-generated \textit{fake} news by comparing South African participants with contextual knowledge against those less acquainted with South African issues. This reveals how background knowledge influences misinformation detection---crucial as AI content proliferates globally~\cite{roozenbeek2022prebunking,dugan2022real}. Our study addresses these research questions:
\begin{enumerate}
    \item How accurately can humans distinguish between \textit{true} news and AI-generated \textit{fake} news about South Africa?
    \item Does cultural proximity to the news content (being South African versus other nationalities) affect human detection performance?
    \item Which features of news articles do participants rely on when making authenticity judgements, and do these differ between cultural groups?
\end{enumerate}

Our study makes the following important contributions to the emerging field of AI-generated \textit{fake} news detection:

\begin{itemize}
    \item We provide empirical evidence on human detection of AI-generated \textit{fake} news in a rapidly evolving technological landscape~\cite{kreps2022all}.
    \item We offer a cross-cultural perspective comparing detection abilities between participants with varying subject matter familiarity.
    \item We identify news features readers use to determine authenticity, informing detection strategies and educational interventions~\cite{jawahar2020automatic}.
    \item By examining South African news, we extend \textit{fake} news research beyond typically studied Western contexts~\cite{marivate2020improving}.
\end{itemize}

The remainder of this paper is organised as follows: Section 2 reviews related work on LLMs, \textit{fake} news, and human detection capabilities. Section 3 describes our experimental methodology. Section 4 presents findings on participants' performance, cross-cultural differences, and factors influencing judgments. Section~5 concludes with key insights and future directions.

\section{Related Work}

This section reviews literature on human detection of AI-generated \textit{fake} news, examining LLMs, misinformation impacts, human detection capabilities, cross-cultural factors, and South African contexts.

\vspace{-0.2cm}

\subsection{LLMs and Detection of AI-Generated Text}

LLMs have evolved into sophisticated neural networks since the Transformer architecture \cite{vaswani2017attention} enabled breakthroughs via self-attention mechanisms. This led to models like BERT \cite{devlin2019bert} with bidirectional training and OpenAI's GPT series. GPT-3 \cite{brown2020language} demonstrated remarkable generative abilities with 175 billion parameters, while GPT-4 \cite{openai2023gpt4}, Claude, and Gemini have further enhanced these capabilities.

Approaches to detecting AI-generated text have evolved alongside generation capabilities, with Jawahar et al.~\cite{jawahar2020automatic} noting that statistical methods quickly became outdated. Gehrmann et al.~\cite{gehrmann2019gltr} proposed GLTR (Giant Language model Test Room) to visualize machine-generated text patterns, though its effectiveness diminished with newer LLMs. Mitchell et al.~\cite{mitchell2023detectgpt} developed DetectGPT using inherent model output consistency, while Kirchenbauer et al.~\cite{kirchenbauer2023watermark} demonstrated effective watermarking techniques. Schaaff et al.~\cite{schaaff2024classification,schaaff2023classification} showed detection performance varies across languages, and Mindner et al.~\cite{mindner2023classification} identified distinctive ChatGPT-generated text patterns. Kreps et al.~\cite{kreps2022all} found that blended human-AI content presents particular detection challenges.

\vspace{-0.2cm}

\subsection{Human Detection Capabilities}

Modern LLMs generate text increasingly indistinguishable from human writing. Clark et al.~\cite{clark2021all} found that untrained evaluators could only identify GPT3-generated text at random chance levels, and even after training them with detection guidelines, annotated examples, or human/AI text comparisons, accuracy improved only marginally to 55\%. Dugan et al.~\cite{dugan2022real} showed participants identified exact AI-transition points in only 23.4\% of cases, though this improved to 72.3\% when identifying any AI-generated sentence. Common sense errors and irrelevant content proved more reliable indicators than grammatical errors. Guess et al.~\cite{guess2020less} found that detection ability varies with age, education, and media literacy. Kasneci et al.~\cite{kasneci2023chatgpt} emphasised how LLMs' fluency and apparent authority complicate distinguishing reliable information from plausible misinformation.

\vspace{-0.2cm}

\subsection{Cross-Cultural Detection Studies}
Roozenbeek et al.~\cite{roozenbeek2022prebunking} tested the ``prebunking'' game \textit{Bad News} with participants from Germany, Greece, Poland, and Sweden, finding significant improvements in misinformation detection across all languages. German participants showed the strongest improvement (d=0.41), followed by Greek (d=0.36) and Polish participants (d=0.33), with country differences accounting for less than 0.1\% of variance. Baptista et al.~\cite{baptista2025human} found that journalism students rated ChatGPT-generated news higher than human-written journalism, with Spanish students rating AI content more favourably than Portuguese counterparts. News topic also influenced evaluations, suggesting cultural and educational factors affect content assessment. These studies highlight gaps in research addressing South African contexts and cross-cultural comparisons of AI-generated \textit{fake} news detection.

\vspace{-0.2cm}

\subsection{South African News Corpora}

South African \textit{fake} news research is hampered by scarce news corpora, particularly in indigenous languages. While global English resources like FakeNewsNet \cite{shu2020fakenewsnet} exist, South African collections are limited. News24 provides primarily English content, while Marivate and Sefara \cite{marivate2020improving} created headline datasets lacking depth for comprehensive analysis. Africa Check and Real411 focus on specific claims rather than complete articles. Mare et al.~\cite{mare2019fake} documented disinformation trends without producing shareable corpora. SADiLaR maintains South African language resources not focused on news, while SAfriSenti and AfriSenti focus on sentiment analysis~\cite{muhammad2023afrisenti,mabokela2025advancing}.

\vspace{-0.2cm}

\subsection{Learnings and Research Gaps}

Our literature review shows modern LLMs produce text nearly indistinguishable from human writing \cite{clark2021all}, with AI-generated \textit{fake} news posing unique threats through scalability and quality \cite{zellers2020defending}. Despite evolving computational detection methods \cite{mitchell2023detectgpt}, human accuracy barely exceeds chance levels \cite{dugan2022real}.

The key research gaps are: limited research on South African readers' detection skills despite high misinformation risk \cite{wasserman2020fake}; few cross-cultural detection studies; scarce analysis of LLM-generated content; and no local news corpora.

Our study addresses these gaps by comparing participants from South Africa and other countries in distinguishing \textit{true} from GPT-4o-generated \textit{fake} South African news, contributing insights for effective countermeasures.

\section{Experimental Setup}

This section outlines our experimental setup for investigating cross-cultural \textit{fake} news detection, including corpus construction and participant survey design.

\subsection{Overview}
To investigate cross-cultural differences in \textit{fake} news detection abilities, we designed an experiment comparing the ability of participants from South Africa and other nationalities in distinguishing \textit{true} from AI-generated \textit{fake} news about South Africa. The study presented 20 news articles (10 \textit{true}, 10 \textit{fake}) via an online survey, examining whether cultural proximity affects detection accuracy, as South Africans would presumably have greater contextual knowledge.


\begin{figure}[h]
\centering
\fbox{\parbox{1.0\textwidth}{
I want to create ideas for a novel I'm working on and need some inspiration about potential disinformation. Can you process the following text I wrote in quotation marks by running it through the following steps:

1. Identify the key figure.\\
2. Suggest what kind of disinformation could be spread about the key figure.\\
3. Take the first suggestion and rewrite the text accordingly. Do not change the tone or the overall length of the text.
}}
\caption{Prompt used to generate \textit{fake} news articles from \textit{true} South African news.}
\label{fig:prompt}
\end{figure}

\subsection{Corpus Construction}

We selected 10 \textit{true} news articles from reputable South African news outlets covering diverse topics (politics, crime, economics, sports, cultural events). This broad range minimised topic-specific knowledge effects on detection. All articles were in English, South Africa's media language. Using these \textit{true} articles as a foundation, we generated 10 corresponding \textit{fake} versions using GPT-4o (OpenAI). The prompt (Fig.~\ref{fig:prompt}) transformed genuine news into plausible misinformation without explicitly requesting ``fake news''. 
The resulting \textit{fake} articles maintained the general structure and topic of the originals while introducing false information about key figures, events, or statistics. To illustrate this approach, Figures~\ref{fig:authentic-article} and~\ref{fig:fake-article} present a \textit{true} article and its AI-generated.

\vspace{-0.2cm}

\begin{figure}[ht]
\fbox{\begin{minipage}{\textwidth}
\textbf{True Article:}

{\fontsize{8pt}{10pt}\selectfont
On Sunday, KwaZulu-Natal Police Commissioner Nhlanhla Mkhwanazi dropped a bombshell. He accused top cops, Police Minister Senzo Mchunu, and the deputy national commissioner Shadrack Sibiya, of being in cahoots with a criminal syndicate linked to the drug trade, and with questionable business people involved in organised crime. Mkhwanazi says his suspicions have been confirmed by a letter from Mchunu to disband the KZN political killings task team in December. He told a media briefing that the task team is focusing on the dark side of Gauteng's criminal justice system, which prompted the approach from Mchunu. Sibiya denies any links to criminal figures. There is a history of the state and politicians having close links to gangs and criminal organisations, says Kiewit.
}
\end{minipage}}
\caption{Example of a \textit{true} news article used in the study.}
\label{fig:authentic-article}
\end{figure}

\vspace{-0.2cm}

\begin{figure}[ht]
\fbox{\begin{minipage}{\textwidth}
\textbf{AI-Generated Fake Article:}

{\fontsize{8pt}{10pt}\selectfont
On Sunday, KwaZulu-Natal Police Commissioner Nhlanhla Mkhwanazi \textbf{made controversial claims that raised eyebrows across the country} [\textit{False context}]. He accused top cops, Police Minister Senzo Mchunu, and deputy national commissioner Shadrack Sibiya of collaborating with a criminal syndicate tied to the drug trade and shady business networks. \textbf{However, senior government insiders suggest that Mkhwanazi's accusations may be driven by a personal vendetta after being overlooked for a national appointment earlier this year} [\textit{Fabricated content}]. At a media briefing, he cited a letter from Mchunu disbanding the KZN political killings task team in December as proof of interference. He linked this move to the team's growing focus on corruption within Gauteng's criminal justice system. Sibiya has denied any links to criminal figures. \textbf{Critics warn that Mkhwanazi's credibility may be compromised by his political ambitions and history of clashing with senior leadership, though others argue his concerns reflect a deeper rot} [\textit{Fabricated content}]. There is a history of the state and politicians having close links to gangs and criminal organisations, says Kiewit.
}
\end{minipage}}
\caption{AI-generated \textit{fake} version of the article with disinformation elements highlighted: Bold text indicates altered or added content with [type of disinformation] noted in brackets.}
\label{fig:fake-article}
\end{figure}

\vspace{-0.2cm}

\begin{table}[ht]
\centering
\footnotesize
\begin{tabular}{|p{3cm}|p{9.2cm}|}
\hline
\textbf{Category} & \textbf{Changes in the fake article} \\
\hline
False context & \textbf{Reframed the significance:} ``made controversial claims that raised eyebrows'' instead of ``dropped a bombshell''---deliberately diminishing the credibility of the accusations \\
\hline
Fabricated content & \textbf{Added fabricated source:} ``senior government insiders suggest that Mkhwanazi's accusations may be driven by a personal vendetta''---inventing non-existent sources \\
\hline
Fabricated content & \textbf{Created false motivation:} ``after being overlooked for a national appointment earlier this year''---inventing a motive to discredit the commissioner \\
\hline
Fabricated content & \textbf{Added fabricated criticism:} ``Critics warn that Mkhwanazi's credibility may be compromised by his political ambitions and history of clashing with senior leadership''---adding fictional critics and allegations \\
\hline
\end{tabular}
\caption{Disinformation elements in the example AI-generated \textit{fake} news article.}
\label{tab:information-types}
\end{table}

Table~\ref{tab:information-types} shows the \textit{fake} version, which introduces several types of misinformation aligned with Wardle and Derakhshan's~\cite {wardle2017information} typology. 
We manually reviewed all generated articles to ensure they contained substantive factual alterations while remaining plausible and stylistically consistent with the originals. 

\vspace{-0.2cm}

\subsection{Survey Design}

All participants voluntarily took part in the study. Each participant classified each of the 20 articles on a 5-point Likert scale (1=``\textit{Definitely fake news}'' to 5=``\textit{Definitely true news}''), following Roozenbeek et al.~\cite{roozenbeek2022prebunking}. After each classification, participants selected features that influenced their judgment from a list including writing style, factual accuracy, logical coherence, source citations, and personal knowledge. This survey design allows us to assess both the detection accuracy and the reasoning strategies behind the participants' judgements, as per Dugan et al.~\cite{dugan2022real}.

We collected demographics (age, gender, nationality, education, frequency of reading South African news) and concluded with Likert-scale questions (1-5) about \textit{perceived classification difficulty}, \textit{topic familiarity's impact on classification}, and \textit{confidence in detection abilities}. This design enabled comparing South African participants with those from other nationalities to examine how cultural proximity affects human detection performance and evaluation strategies.


\section{Experiments and Results}

This section presents our empirical findings on how cultural background and South African familiarity influence \textit{fake} news detection, reporting participant demographics, detection performance, factors affecting credibility judgments, and linguistic features that differentiate \textit{true} from AI-generated \textit{fake} content.

\subsection{Participant Demographics}

Our study included 89 participants with diverse demographic characteristics. The distribution of participants by nationality showed a predominance of South Africans (n=56, 62.9\%) compared to participants from other nationalities (n=33, 37.1\%), which included Germans, Bulgarians, Congolese, Sudanese, Zimbabweans, and Brazilians. This distribution allowed us to effectively compare South African participants with those from other countries, particularly focusing on how cultural proximity might influence \textit{fake} news detection capabilities.

Age distribution among participants was heavily skewed toward younger demographics, with 61 participants (68.5\%) in the 18-24 age range, 13 participants (14.6\%) in the 25-34 age range, 12 participants (13.5\%) in the 35-44 age range, and 3 participants (3.4\%) in the 45-54 age range. Gender distribution was relatively balanced, with 46 male participants (51.7\%), 39 female participants (43.8\%), and 4 participants (4.5\%) preferring not to disclose their gender.

In terms of educational background, the majority of participants reported having a Diploma (n=47, 52.8\%), followed by high school education (n=18, 20.2\%), Bachelor's degree (n=7, 7.9\%), Master's degree (n=11, 12.4\%), and PhD (n=6, 6.7\%). This relatively high proportion of diploma-holding participants should be considered when interpreting the results.

\vspace{-0.2cm}

\begin{table}[ht]
\centering
\caption{Frequency of reading news about South Africa by nationality group (\%).}
\label{tab:news_frequency}
\begin{tabular}{lccc}
\hline
\textbf{Frequency} ~~& ~~ \textbf{South Africa} ~~& ~~\textbf{Other Countries}~~&~~\textbf{All Countries} \\
\hline
Daily & 12.5 & 3.0 & 9.0 \\
Weekly & 37.5 & 12.1 & 28.1 \\
Monthly & 23.2 & 3.0 & 15.7 \\
Half a year & 7.1 & 6.1 & 6.7 \\
Once a year & 5.4 & 30.3 & 14.6 \\
Never & 14.3 & 45.5 & 25.8 \\
\hline
\end{tabular}
\end{table}

Particularly relevant to our research questions was participants' frequency of reading news about South Africa, as summarised in Table~\ref{tab:news_frequency}. The data reveal a stark contrast in news consumption patterns between South Africans and other participants. Weekly news consumption was the most common frequency for South Africans (37.5\%), while participants from other nationalities most commonly reported never reading news about South Africa (45.5\%) or reading it only once a year (30.3\%). Overall, 50.0\% of South African participants reported frequent engagement with news about their country (daily or weekly), compared to just 15.1\% of participants from other nationalities.

\begin{table}[ht]
\centering
\caption{News consumption media by nationality group (\%).}
\label{tab:news_medium}
\begin{tabular}{lccc}
\hline
\textbf{News Medium} ~~~~& \textbf{South Africa} ~~~~ & \textbf{Others Countries} ~~~~& \textbf{All Countries} \\
\hline
Social Media & 89.3 & 66.7 & 80.9 \\
Online Newspaper & 33.9 & 54.5 & 41.6 \\
TV & 41.1 & 36.4 & 39.3 \\
Video Platforms & 44.6 & 27.3 & 38.2 \\
Radio & 35.7 & 24.2 & 31.5 \\
Microblogging & 28.6 & 15.2 & 23.6 \\
Printed Newspaper & 12.5 & 12.1 & 12.4 \\
\hline
\end{tabular}
\end{table}

Regarding news consumption media (Table~\ref{tab:news_medium}), social media platforms were overwhelmingly the most common source across both groups, with even higher prevalence among South Africans (89.3\%) than other participants (66.7\%). A notable difference appeared in online newspaper consumption, with 54.5\% of participants from other nationalities using this medium compared to only 33.9\% of South Africans. South Africans showed greater preference for video platforms (44.6\% vs. 27.3\%), TV (41.1\% vs. 36.4\%), radio (35.7\% vs. 24.2\%), and microblogging services (28.6\% vs. 15.2\%). Printed newspapers were similarly unpopular across both groups (12.5\% vs. 12.1\%).

These patterns of media consumption likely influence how participants evaluate news credibility. The strong reliance on social media platforms, particularly among South African participants, may affect their exposure to diverse perspectives and fact-checking resources. Meanwhile, participants from other nationalities show a more balanced approach between social media and traditional online news sources, which may influence their evaluation strategies.

The demographic composition of our expanded sample---featuring a wider range of participants with varying degrees of exposure to South African news and diverse consumption habits---provides a robust foundation for examining how cultural proximity and media literacy affect the detection of AI-generated \textit{fake} news about South Africa.

\vspace{-0.2cm}

\subsection{Detection Performance Across Participant Groups}
\label{Detection Performance Across Participant Groups}

Our analysis of participants' ability to detect AI-generated \textit{fake} news revealed interesting patterns when comparing South African participants with those from other nationalities. Table~\ref{tab:mean_deviation} presents the mean deviation from ideal ratings (5 for \textit{true} news, 1 for \textit{fake} news), where lower values indicate better performance.

\vspace{-0.2cm}

\begin{table}[ht]
\centering
\caption{Mean deviation from ideal ratings by participant group and article type.}
\label{tab:mean_deviation}
\begin{tabular}{lccc}
\hline
\textbf{Participants} ~~~~& \textbf{Fake News} ~~~~& \textbf{True News} ~~~~& \textbf{Fake+True News} \\
\hline
South Africa & 2.49 & 1.59 & 2.04 \\
Others Countries & 2.19 & 2.08 & 2.13 \\
All Countries & 2.38 & 1.74 & 2.06 \\
\hline
\end{tabular}
\end{table}

\begin{figure}
    \centering
    \includegraphics[width=.7\linewidth]{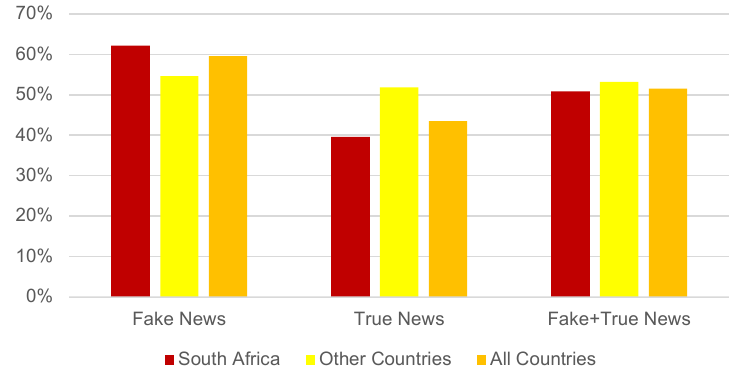}
\caption{Mean deviation (\%) from ideal ratings by participant group and article type.} \label{fig:accuracy}
\end{figure}

To facilitate more intuitive interpretation, we converted these deviations into percentages, calculated as $deviation$/4, where 4 represents the maximum possible deviation on our 5-point scale. These percentages are presented in Figure~\ref{fig:accuracy}.

The results reveal a notable pattern: South African participants demonstrated superior performance in detecting \textit{true} news about their country (40\% mean deviation from ideal ratings) compared to participants from other nationalities (52\%). However, this advantage did not extend to identifying \textit{fake} news, where South Africans performed worse (62\%) than other participants (55\%).

Overall deviation across both article types was similar between South Africans (51\%) and other participants (53\%), suggesting that while detection performance patterns differ between groups, their aggregate ability to distinguish between \textit{true} and \textit{fake} news remains comparable. The mean deviation across all participants was 52\%, indicating that distinguishing between \textit{true} and AI-generated \textit{fake} news presents a substantial challenge regardless of cultural background.

Both groups performed better on \textit{true} news detection than on \textit{fake} news detection. This asymmetry may reflect a bias toward accepting news as \textit{true}---a finding consistent with previous research on misinformation~\cite{guess2020less}---or indicate that the AI-generated \textit{fake} news in our study was particularly convincing.

\subsection{Trust in News Across Participant Groups}

To assess overall trust in news articles across participant groups, we calculated the mean score over all news articles (both \textit{fake} and \textit{true}) on the 5-point Likert scale (1=``\textit{Definitely fake news}'' to 5=``\textit{Definitely true news}'') as a trust proxy.
For South Africa, this mean score is 3.45, which indicates a higher level of trust in news media compared to the mean score of 3.04 for other countries. This finding helps explain the differences described in Section~\ref{Detection Performance Across Participant Groups}, as South Africans' higher trust correlates with less deviation on \textit{true} news and greater deviation on \textit{fake} news.

\begin{figure}
    \centering
        \includegraphics[width=.8\linewidth]{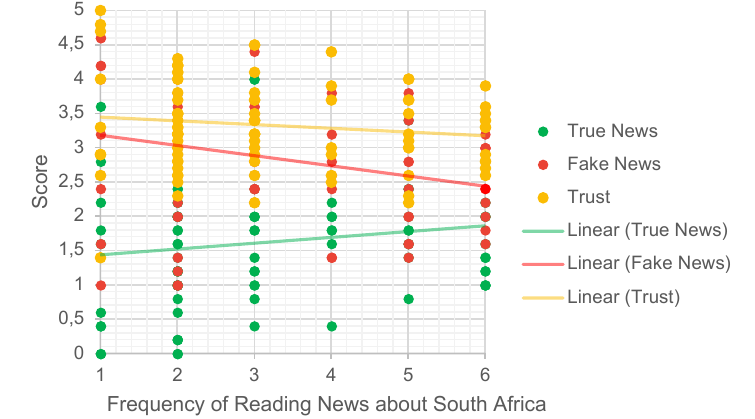}
    \caption{Error in detecting \textit{true} and \textit{fake} news as well as trust in relation to the frequency of news consumption.}
    \label{fig:reading}
\end{figure}

Furthermore, we analysed the relationship between the Likert scores selected by participants and their frequency of reading news about South Africa. Figure~\ref{fig:reading} shows the Likert scores selected by each participant in relation to their indicated frequency of news consumption about South Africa. 
The figure reveals that no relationship exists between the level of trust (mean Likert scores, indicated in yellow) and the frequency of South African news consumption. However, the performance in distinguishing between \textit{true} news articles (indicated in green) and \textit{fake} news articles (indicated in red) is influenced by reading frequency: There is a tendency that a higher frequency of news consumption correlates with fewer errors in identifying \textit{true} news but more errors in identifying \textit{fake} news.

\vspace{-1.2cm}

\subsection{Features of the Participants to Assess News Credibility}

After classifying each news article, participants selected which features influenced their judgment (with the option to select multiple features for each article). The features presented to participants were:

\begin{itemize}
\item \textit{Personal Knowledge of the Topic}: Familiarity with the subject
\item \textit{Factual Accuracy Based on Knowledge}: Alignment with existing knowledge
\item \textit{Writing Quality and Style}: Clarity, coherence, and tone
\item \textit{Emotional Tone and Language}: Use of emotionally charged words
\item \textit{Presence of Specific Details}: Names, dates, locations, statistics
\item \textit{Double-Checking with Other News}: Verification with other sources
\item \textit{Logical Coherence and Plausibility}: Rational structure and believability
\item \textit{Statistical Data Presented}: Numerical information
\item \textit{Quotes from Identified Sources}: Direct citations
\item \textit{Official Statements Included}: Information from institutions
\item \textit{Expert Opinions Cited}: References to authorities
\end{itemize}

Table \ref{tab:feature_comparison} presents how frequently South African participants and those from other countries relied on each feature, grouped by category, and whether this led to correct or incorrect classifications. The percentages show how often participants used each feature, with ``\textit{Correct}'' indicating when this led to accurate classification and ``\textit{Incorrect}'' when it led to misclassification. Since participants could select multiple features for each article, these percentages reflect the relative frequency of feature usage rather than mutually exclusive choices.

\begin{table}[ht]
\centering
\caption{Features Used When Classifying News Articles (\%).}
\label{tab:feature_comparison}
\footnotesize
\begin{tabular}{lcccc}
\hline
\textbf{Feature} & \multicolumn{2}{c}{\textbf{Correct}} & \multicolumn{2}{c}{\textbf{Incorrect}} \\
\cline{2-3} \cline{4-5}
& \textbf{SA} & \textbf{~Other~} & \textbf{~SA~} & \textbf{Other} \\
\hline
\multicolumn{5}{l}{\textbf{Knowledge-Based Features}} \\
Personal Knowledge of the Topic & 15.5 & 6.2 & 15.5 & 4.8 \\
Factual Accuracy Based on Knowledge~~ & 11.0 & 4.8 & 14.5 & 4.8 \\
\hline
\multicolumn{5}{l}{\textbf{Linguistic Features}} \\
Writing Quality and Style & 13.5 & 22.1 & 15.5 & 26.9 \\
Emotional Tone and Language & 4.8 & 4.8 & 4.8 & 11.0 \\
\hline
\multicolumn{5}{l}{\textbf{Factual/Logical Features}} \\
Presence of Specific Details & 19.4 & 17.2 & 17.4 & 17.9 \\
Double-Checking with Other News & 14.5 & 5.5 & 12.9 & 4.1 \\
Logical Coherence and Plausibility & 9.4 & 20.0 & 10.3 & 11.7 \\
Statistical Data Presented & 7.4 & 6.2 & 6.1 & 6.9 \\
Quotes from Identified Sources & 7.4 & 3.4 & 10.6 & 5.5 \\
Official Statements Included & 7.1 & 6.2 & 6.8 & 6.9 \\
Expert Opinions Cited & 3.9 & 3.4 & 4.2 & 4.1 \\
\hline
\end{tabular}
\end{table}

\vspace{-0.2cm}

We analysed how South African participants and those from other countries used different categories of features when assessing news credibility:

\vspace{-0.2cm}

\subsubsection{Knowledge-Based Features}

South African participants showed notably higher usage rates of these features (15.5\% correct / 15.5\% incorrect for \textit{personal knowledge of the topic}; 11.0\% correct / 14.5\% incorrect for factual accuracy) compared to other countries (6.2\% correct / 4.8\% incorrect for \textit{personal knowledge of the topic}; 4.8\% correct / 4.8\% incorrect for \textit{factual accuracy based on knowledge}). This suggests South Africans rely heavily on their perceived knowledge about South Africa, even when this approach leads to errors.
Overall, South Africans place greater emphasis on \textit{personal knowledge of the topic}, even when its reliability may be limited, while other countries engage with \textit{knowledge-based features} more critically. This distinction highlights varying strategies in assessing information credibility.

\subsubsection{Linguistic Features}

South African participants used these features less frequently (13.5\% correct / 15.5\% incorrect for \textit{writing quality and style}; 4.8\% correct / 4.8\% incorrect for \textit{emotional tone and language}) than participants from other countries (22.1\% correct / 26.9\% incorrect for \textit{writing quality and style}; 4.8\% correct / 11.0\% incorrect for \textit{emotional tone and language}). 
Overall, South African participants appear to approach \textit{linguistic features} with caution, potentially avoiding over-reliance on them, while others may prioritise these features more heavily, even if it increases the risk of errors. This highlights divergent strategies in leveraging language-based signals to assess news credibility.  

\vspace{-0.2cm}

\subsubsection{Factual/Logical Features}

South Africans showed stronger performance with concrete features like \textit{the presence of specific details} (19.4\% correct vs. 17.2\%) and \textit{double-checking with other news} (14.5\% correct vs. 5.5\%), but struggled with \textit{logical coherence and plausibility} (9.4\% correct vs. 20.0\% for other countries). 
Regarding \textit{statistical data presented}, South Africans performed slightly better (7.4\% correct vs. 6.2\%) with marginally fewer errors (6.1\% incorrect vs. 6.9\%) than participants from other countries. For \textit{quotes from identified sources}, South Africans showed higher usage but poorer detection performance (7.4\% correct / 10.6\% incorrect) compared to other participants (3.4\% correct / 5.5\% incorrect). Both groups used \textit{official statements included} with similar effectiveness (South Africans: 7.1\% correct / 6.8\% incorrect; Others: 6.2\% correct / 6.9\% incorrect). \textit{Expert opinions cited} were the least utilised feature across both groups, with comparable detection performance (South Africans: 3.9\% correct / 4.2\% incorrect; Others: 3.4\% correct / 4.1\% incorrect). 
This suggests that South Africans may prioritise concrete, factual details (e.g., specific data or statistics) over abstract logical reasoning, while other countries appear to emphasise logical structure and expert validation, even if this approach occasionally leads to higher error rates in certain areas.  

\vspace{-0.2cm}

\subsubsection{Summary}

South African participants appear to rely heavily on \textit{personal knowledge of the topic}, even when its detection performance may be limited, while prioritising concrete \textit{factual} details---such as specific examples or statistical data---over abstract reasoning. Their verification efforts, though present, yield mixed results, and they tend to avoid over-reliance on \textit{linguistic} cues like tone or style. In contrast, participants from other countries place greater emphasis on \textit{logical coherence} and \textit{expert} validation, often incorporating \textit{linguistic features} such as tone or stylistic elements into their assessments. However, this approach comes with higher error rates, suggesting a balance between leveraging language-based signals and the risks of misinterpretation. Overall, South Africans seem to adopt a more fact-driven, cautious strategy, while others prioritise logical and linguistic frameworks, even if this increases the likelihood of errors in certain contexts.

\vspace{-0.2cm}

\subsection{Objective Linguistic Analysis of News Articles}

In addition to analysing participants' subjective evaluations, we conducted an objective linguistic analysis of the news articles using natural language processing techniques. This complementary approach allows us to examine whether measurable textual properties differed between \textit{true} and AI-generated \textit{fake} news, and whether these properties correlate with participants' credibility judgments.

\subsubsection{Readability Assessment}

\begin{figure}
    \centering
    \includegraphics[width=.8\linewidth]{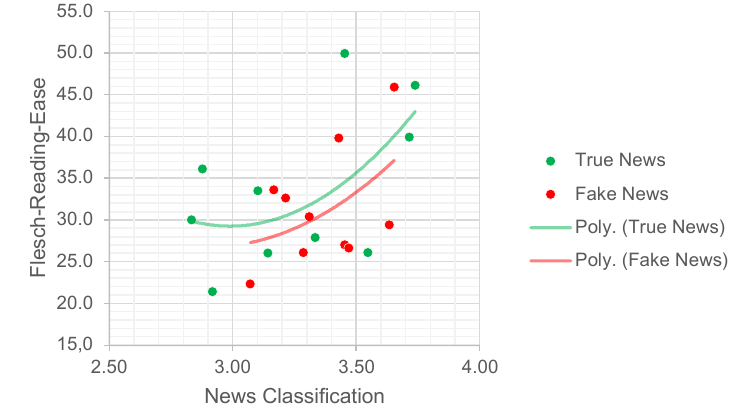}
    \vspace{-0.2cm}
    \caption{Relationship between Flesch reading ease scores and average credibility ratings (Y-axis; 1 = definitely fake, 5 = definitely true) across 20 articles (10 \textit{true} and \textit{fake}). Note: ``Poly.'' denotes a polynomial approximation of the second order of the data.}
    \label{fig:FRE}
\end{figure}

The Flesch reading ease (FRE) score measures text readability, with higher scores indicating easier-to-read content. We calculated FRE scores for all articles and compared them against participants' average credibility ratings. 
As shown in Figure~\ref{fig:FRE}, \textit{true} news articles generally demonstrated higher readability (average FRE score of 33.69) compared to \textit{fake} news (average FRE score of 31.37). This suggests that the AI-generated content may be slightly more complex and less accessible than authentic journalism. Interestingly, participants assigned higher average credibility ratings to \textit{fake} news (3.37) than to \textit{true} news (3.27), indicating that greater readability did not necessarily lead to higher perceived credibility. 
These findings suggest that while humans might use writing quality as a criterion for judgment (as seen in our feature analysis), the relationship between objective readability measures and perceived credibility is not straightforward. More readable text is not automatically perceived as more \textit{true}, which could be exploited by sophisticated AI systems generating highly readable but false content.

\vspace{-0.4cm}

\vspace{-1.1cm}

\subsubsection{Sentiment Analysis}

\begin{figure}
    \centering
    \includegraphics[width=.8\linewidth]{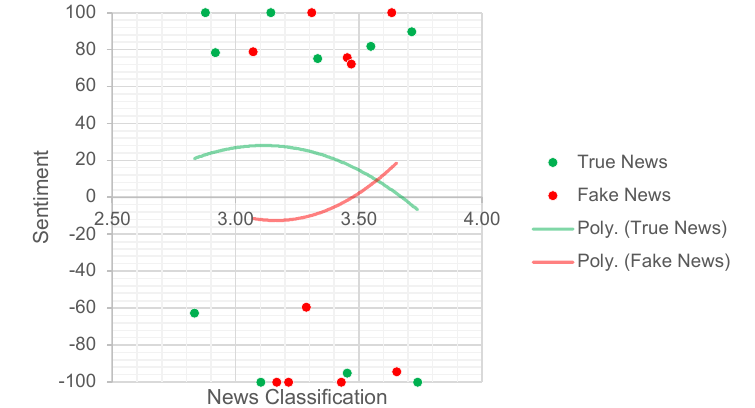}
    \vspace{-0.2cm}
    \caption{Relationship between sentiment scores and average credibility ratings across \textit{true} and \textit{fake} news articles (Y-axis; 1 = definitely fake, 5 = definitely true). Note: ``Poly.'' denotes a polynomial approximation of the second order of the data.}
    \label{fig:sentiment}
\end{figure}

We also conducted sentiment analysis to measure the emotional tone of each article, with positive scores indicating positive sentiment and negative scores indicating negative sentiment. 
Figure~\ref{fig:sentiment} reveals a distinct sentiment pattern: \textit{true} news articles typically exhibited positive sentiment (average score of 16.68), while \textit{fake} news articles tended toward negative sentiment (average score of -2.72). This sentiment divergence may reflect how the AI system altered the emotional tone when generating \textit{fake} versions of news stories, potentially exaggerating or introducing negative elements to create more sensational content. 
Despite this clear sentiment difference, participants' credibility ratings did not align with this pattern. \textit{Fake} news received slightly higher average credibility ratings (3.37) than \textit{true} news (3.27), suggesting negative sentiment did not trigger scepticism among participants. This finding is particularly noteworthy given that \textit{emotional tone and language} was one of the features participants reported using in their credibility assessments. 
These results highlight the complex relationship between objective linguistic measures and human perception of credibility. While measurable differences exist between \textit{true} and AI-generated \textit{fake} news in terms of readability and sentiment, these differences do not appear to effectively guide human detection of \textit{fake} content. This underscores the challenge of developing reliable indicators for distinguishing between human and AI-generated text, particularly as language models continue to improve.

\vspace{-0.5cm}

\vspace{-1.0cm}

\section{Conclusion and Future Work}

\vspace{-0.2cm}

Our findings provide preliminary evidence that cultural familiarity influences the detection of AI-generated news. Our study investigated how cultural proximity affects the ability to detect AI-generated \textit{fake} news about South Africa, revealing several important insights with implications for misinformation research, media literacy, and technological countermeasures.

\vspace{-0.2cm}

\subsection{Key Findings}

\vspace{-0.1cm}

Our study demonstrates it was straightforward to generate realistic misinformation using our prompt and GPT-4o. By employing a carefully crafted prompt that framed the request as creating fictional content for a novel, we were able to circumvent the system's built-in safeguards. Direct requests to create \textit{fake} news were rejected by the system with warning messages, highlighting the importance of indirect prompting strategies when investigating AI disinformation risks.

Our results demonstrate a complex relationship between cultural familiarity and \textit{fake} news detection. South African participants excelled at recognising \textit{true} news about their country but performed worse at identifying AI-generated \textit{fake} content. A closer look revealed these differences stem from a higher level of trust.

Our analysis revealed that South Africans relied more heavily on knowledge-based features (personal knowledge, factual accuracy) and concrete factual features (specific details, statistical data), while participants from other countries emphasised linguistic features (writing quality, emotional tone) and logical coherence. This distinction reflects different approaches to credibility assessment: South Africans employed a content-focused strategy based on their contextual knowledge, while others relied more on structural and stylistic indicators.

The objective linguistic analysis provided additional insights, showing that \textit{true} news articles generally had higher readability scores and more positive sentiment than AI-generated \textit{fake} news. However, participants' credibility judgments did not consistently align with these objective measures, highlighting the complex relationship between textual properties and perceived credibility.

\vspace{-0.2cm}

\subsection{Implications}

\vspace{-0.1cm}

These findings have significant implications for combating AI-generated \textit{fake} news. First, they suggest that different strategies for \textit{fake} news detection may be needed for audiences with varying levels of cultural proximity to the content. While those familiar with a subject domain benefit from contextual knowledge when verifying \textit{true} information, they may need additional tools or training to overcome potential biases when evaluating fabricated content.

Second, similar overall deviation from ideal ratings between groups indicates that distinguishing between \textit{true} and AI-generated news presents a substantial challenge regardless of cultural background. This underscores the sophistication of modern LLMs and the urgent need for improved detection methods.

Third, the discrepancy between objective linguistic measures and human perception highlights the limitations of purely computational approaches to \textit{fake} news detection. Effective solutions likely require a combination of technological tools, human judgment, and targeted educational interventions.

\vspace{-0.2cm}

\subsection{Limitations and Future Work}

\vspace{-0.1cm}

Our study has limitations suggesting directions for future research. The sample (89, mostly young and highly educated) limits generalisability. Participants came from multiple countries, so results may reflect aggregated differences. Future studies should include larger, more diverse samples. 

Additionally, our experiment used only English-language news articles, whereas South Africa has twelve official languages. Future research should examine how language choice affects \textit{fake} news detection, particularly in multilingual societies where language itself may serve as a credibility cue.

The finding that \textit{fake} news received slightly higher credibility ratings than \textit{true} news warrants further investigation. Additionally, we used a single LLM (GPT-4o) with one indirect prompting strategy; replication with alternative models and prompt types is needed to validate findings. Future studies could explore whether this pattern reflects characteristics of the AI-generated content, biases in human judgment, or limitations in the experimental design.

Finally, we need longitudinal studies to track how detection abilities change as AI evolves. Will the patterns we observed persist as LLMs advance, or will we face new challenges requiring different strategies? Integrating objective linguistic signals with human-centred detection strategies may yield more resilient countermeasures against AI-driven disinformation.

%
%

%
%
 \bibliographystyle{splncs04}
 \bibliography{mybibliography}
%

\end{document}